\documentclass[prc,amsmath,twocolumn,showkeys,showpacs,superscriptaddress]{revtex4}

\usepackage{graphicx,color}
\usepackage{amssymb}
\usepackage{enumerate}
\usepackage{verbatim}
\usepackage{natbib}

\begin{document}
  \newcommand {\nc} {\newcommand}
  \nc {\Sec} [1] {Sec.~\ref{#1}}
  \nc {\IR} [1] {\textcolor{red}{#1}} 

\title{Testing the Perey Effect}

\author{L.~J.~Titus}
\affiliation{National Superconducting Cyclotron Laboratory, Michigan State University, East Lansing, MI 48824, USA}
\affiliation{Department of Physics and Astronomy, Michigan State University, East Lansing, MI 48824-1321}
\author{F.~M.~Nunes}
\affiliation{National Superconducting Cyclotron Laboratory, Michigan State University, East Lansing, MI 48824, USA}
\affiliation{Department of Physics and Astronomy, Michigan State University, East Lansing, MI 48824-1321}

\date{\today}


\begin{abstract}
{\bf Background:} The effects of non-local potentials have historically been approximately included by applying a correction 
factor to the solution of the corresponding equation for the local equivalent interaction. This is usually referred to as the 
Perey correction factor.
{\bf Purpose:} In this work we investigate the validity of the Perey correction factor for single-channel bound and scattering 
states, as well as in transfer $(p,d)$ cross sections.
{\bf Method:}  We solve the scattering and bound state equations  for non-local interactions of the Perey-Buck type, through an 
iterative method. Using the distorted wave Born approximation, we  construct the T-matrix for $(p,d)$  on $^{17}$O, $^{41}$Ca, 
$^{49}$Ca, $^{127}$Sn, $^{133}$Sn, and $^{209}$Pb at $20$ and $50$ MeV. 
{\bf Results:}  We found that for bound states, the Perey corrected wave function resulting from the local equation agreed well 
with that from  the non-local equation in the interior region, but discrepancies were found in the surface and peripheral regions. 
Overall, the Perey correction factor was adequate for scattering states, with the exception of a few partial waves corresponding 
to the grazing impact parameters. These differences proved to be important for transfer reactions.
{\bf Conclusions:}  The Perey correction factor does offer an improvement over taking a direct local equivalent solution. 
However, if the desired accuracy is to be better than 10\%, the exact solution of the non-local equation should be pursued.
\end{abstract}

\pacs{21.10.Jx, 24.10.Ht, 25.40.Cm, 25.45.Hi}

\keywords{elastic scattering, bound states, transfer reactions, non-local optical potentials, Perey effect}

\maketitle
\section{Introduction}

Optical potentials are a common ingredient in reaction studies. The optical model potential for nucleon-nucleus scattering has 
long been established as non-local \cite{Bell1959_PRL}. While the vast majority of optical potentials found in the literature are 
assumed to be local and strongly energy dependent (i.e. \cite{Koning2003_NPA,Varner1991_PR}), attempts have been made to 
determine  non-local optical potentials \cite{Perey1962_NP,Giannini1976_AP}. Most often, optical potentials are phenomenologically 
based, fitted to elastic scattering. However, elastic scattering is predominantly an asymptotic process, so the  short range 
details of the interaction cannot be uniquely constrained, particularly short-ranged non-local properties.

The predominant sources of non-locality in the effective nucleon-nucleus interaction arise from anti-symmetrization 
\cite{Canton2005_PRL,Fraser2008_EPJ,Rawitscher1994_PRC}, multiple scattering \cite{Kerman1959_AP,Crespo1996_PRC}, and channel couplings 
\cite{Feshbach1958_AP,Feshbach1962_AP}.  Jeukenne {\it et al.} developed a semi-microscopic formulation for the nucleon-nucleus 
optical potential based on the self-energy \cite{Jeukenne1974_PRC,Jeukenne1977_PRC} . This approach can provide good 
fits to elastic scattering data \cite{Bauge1998_PRC,Bauge2001_PRC}. Another approach consists on defining  an optical potential 
based on the coupling of all inelastic channels to the elastic channel (the Feshbach channel coupling approach).
These non-localities have  been studied by Rawitscher \cite{Rawitscher1987_NPA}. In either approaches, the resulting optical 
potential should be non-local, as well as energy dependent. 

The precise form of the non-local potential is currently not known, and a number of forms have been 
considered (e.g. the parity-dependent potential of Cooper {\it et al.} \cite{Cooper1996_PRC} and the velocity dependent potential 
\cite{Kisslinger1955_PR,Zureikat2013_NPA}). In this work we consider the Perey-Buck form for the non-local potential 
\cite{Perey1962_NP,Frahn1957_NC}, which has energy and mass-independent potential parameters, and consists of a Woods-Saxon 
multiplied by a Gaussian which introduces the range of the non-locality. In \cite{Deltuva2009_PRC}, the effect of non-locality 
of the Perey-Buck form is considered in deuteron induced reactions. Cross sections for various channels obtained when the 
nucleon-nucleus optical potentials are local CH89 \cite{Varner1991_PR} differ considerably from those obtained using the 
non-local GR76 \cite{Giannini1976_AP}. Since both CH89 and GR76 are global potentials with different fitting protocols, it is 
unclear whether the differences seen in \cite{Deltuva2009_PRC} arise from the fact that these potentials are not phase 
equivalent, or genuinely from non-local effects. This deserves closer inspection. 

Accounting for the non-locality through the energy dependence, as done for the local potentials, is known to be insufficient. 
One key feature of a non-local potential is that it reduces the amplitude of the wave function in the nuclear interior compared 
to the wave function from an equivalent local potential \cite{Austern1965_PR,Fiedeldey1966_NP}, the so-called Perey effect. 
Shortly after the introduction of the Perey-Buck potential, Austern studied the wave functions of non-local potentials and 
demonstrated the Perey effect in one-dimension \cite{Austern1965_PR}. Subsequently, Fiedeldey did a similar study but in the 
three-dimensional case \cite{Fiedeldey1966_NP}. Using a different method, Austern presented a way to relate wave functions obtained from 
non-local and local potentials in the three-dimensional case \cite{Austern1970_Book}. 
Since then, non-local calculations have been avoided by using this Perey Correction Factor (PCF). 

Recently, Timofeyuk and Johnson \cite{Timofeyuk2013_PRL,Timofeyuk2013_PRC} studied the effects of including an energy-independent 
non-local potential in $(d,p)$ reactions within the Adiabatic Distorted Wave Approximation (ADWA) \cite{Johnson1974_NPA}. 
Non-locality was included approximately through expansions to construct a local equivalent potential and solving the corresponding 
local Schr\"odinger's equation. They found that a Perey-Buck type non-locality can be effectively included 
in $(d,p)$ through a very significant energy shift in the evaluation of the local optical potentials to be used in constructing the 
deuteron distorted waves. This can impact cross sections dramatically, and calls for further investigations.

In this work, we determine the importance of non-local effects in the various components of a nuclear reaction process, 
and assess the validity of the PCF by studying a wide range of reactions, including neutron states bound to
$^{16}$O, $^{40}$Ca, $^{48}$Ca, $^{126}$Sn, $^{132}$Sn, and $^{208}$Pb, and  $(p,p)$ and $(p,d)$ on
$^{17}$O, $^{41}$Ca, $^{49}$Ca, $^{127}$Sn, $^{133}$Sn, and $^{209}$Pb at $20$ and $50$ MeV. 

The paper is organized in the following way. In \Sec{theory} we briefly describe the necessary theory. Numerical details can be found in \Sec{numerical}. The results are presented 
in \Sec{results}, starting with a discussion of local equivalent potentials in \Sec{LEP} and of 
approximate local equivalent potentials in \Sec{approximate_LEP}. We consider the effects of non-localities on scattering wave 
functions and ways to correct for non-localities in \Sec{WFs}. The effects of non-localities on bound state wave functions are 
presented in \Sec{Bound}. We then explore the effects of non-localities on transfer cross sections in \Sec{Transfer}. 
We discuss the connection of this work with other relevant studies in \Sec{discussion}. 
Finally, in \Sec{conclusions}, conclusions are drawn.

\section{Theoretical considerations}
\label{theory}

Let us consider a nucleon scattering off a composite nucleus. The effective interaction between the nucleon and the nucleus is a 
non-local optical potential. In this case, the two-body Schr\"odinger equation takes the form

\begin{equation}\label{3D-NLeqn}
\frac{\hbar^2}{2\mu}\nabla^2 \Psi(\textbf{r})+E\Psi(\textbf{r})=U_o(\textbf{r})\Psi(\textbf{r})+\int U^{NL}(\textbf{r},\textbf{r}')\Psi(\textbf{r}')d\textbf{r}'
\end{equation}

\noindent where $\mu$ is the reduced mass of the nucleon-nucleus system, $E$ is the energy in the center of mass, 
$U_o(\textbf{r})$ is the local part of the potential, and $\Psi({\bf r})$ is the scattering wave function. A particular form of 
the non-local potential introduced by Frahn and Lemmer \cite{Frahn1957_NC} is

\begin{equation}\label{FrahnLemmer}
U^{NL}(\textbf{r},\textbf{r}')=U^{NL}_{WS}\left(\left|\frac{\textbf{r}+\textbf{r}'}{2} \right| \right)\frac{\exp\left(-\left|\frac{\textbf{r}-\textbf{r}'}{\beta} \right|^2\right)}{\pi^{3/2}\beta^3},
\end{equation}

\noindent where, $\beta$ is the range of the non-locality, and typically takes on a value of $\approx 0.85$ fm. 
In this work, $U^{NL}_{WS}$ is of a Woods-Saxon form of the variable $\tfrac{1}{2}|\textbf{r}+\textbf{r}'|$. 

This type of potential was further investigated by Perey and Buck \cite{Perey1962_NP}. Making the approximation 
$|\textbf{r}+\textbf{r}'| \approx (r+r')$ in $U^{NL}_{WS}$ allows for an analytic partial wave decomposition resulting in the 
partial wave equation

\begin{eqnarray}\label{NLeqn}
\frac{\hbar^2}{2\mu}\left[\frac{d^2}{dr^2}\right.&-&\left.\frac{\ell(\ell+1)}{r^2} \right]\psi^{NL}_\ell(r)+E\psi^{NL}_\ell(r) \nonumber \\
&=&U_o(r)\psi^{NL}_\ell(r)+\int g_\ell(r,r')\psi^{NL}_\ell(r') dr'.
\end{eqnarray}

\noindent Here, the kernel is explicitly given by:

\begin{equation}
g_\ell(r,r')=\frac{2i^\ell z}{\pi^{\frac{1}{2}}\beta}j_\ell(-iz)\exp \left(-\frac{r^2+r'^2}{\beta^2} \right)U^{NL}_{WS}\left(\frac{1}{2}(r+r') \right),
\end{equation}

\noindent where $j_\ell$ are spherical Bessel functions, and $z=2rr'/\beta^2$. In our study, we assume the spin-orbit and Coulomb 
potentials are local, and therefore, $U_o(r)=V_{so}(r)+V_{\textrm{coul}}(r)$. 

For a non-local potential of the Perey-Buck type, the depths of an approximate local equivalent potential can be found from the 
relations \cite{Perey1962_NP}

\begin{eqnarray}\label{NLtoLoc}
V^{NL}_v&=&V^{Loc}_v\exp\left[\frac{\mu\beta^2}{2\hbar^2}\left(E-V_c+V^{Loc}_v \right) \right] \nonumber \\
W^{NL}_d&=&W^{Loc}_d\exp\left[\frac{\mu\beta^2}{2\hbar^2}\left(E-V_c+V^{Loc}_v \right) \right].
\end{eqnarray}

\noindent Here, $V_{v}$ and $W_{d}$ are the depths of the real volume and imaginary surface terms in the Woods-Saxon potential, respectively, 
and are positive constants. $E$ is the center of mass energy, and $V_c$ is the Coulomb potential at the origin for a solid 
uniformly charged sphere with radius $R_c=r_cA^{1/3}$. 
Notice that even though the non-local potential is energy-independent, the transformed local depths are energy-dependent, 
which is a common feature of local global optical potentials.

Through use of Eq.(\ref{NLtoLoc}) and fits to neutron elastic scattering data on $^{208}$Pb at low energies, the Perey-Buck non-local 
potential was determined: the corresponding parameters are given in the first column of Table \ref{tab:Potential_Parameters}. The parameters in the 
Perey-Buck potential are both energy and mass-independent.

For a given non-local potential, a local equivalent potential can often be found. However, in the nuclear interior, the wave 
function resulting from using a non-local potential is reduced compared to the wave function resulting from using a local 
equivalent potential. This phenomenon is known as the Perey effect \cite{Austern1965_PR}. Correcting for the reduced 
amplitude is done via the PCF:

\begin{equation}\label{CorrectionFactor}
F(r)= \left[ 1-\frac{\mu \beta^2}{2 \hbar^2}\left(U^{LE}(r)-U_o(r) \right) \right]^{-1/2}.
\end{equation}

\noindent Note that here, $U^{LE}(r)$ is the local equivalent potential. As we required the spin-orbit and 
Coulomb terms to be identical in the local and non-local potentials, these terms in $U^{LE}$ exactly cancel $U_o$. 
Since $F(r)\rightarrow 1$ as $r \rightarrow \infty$, the correction factor Eq.(\ref{CorrectionFactor}) only affects the magnitude 
of the wave function within the range of the nuclear interaction. A derivation of Eq.(\ref{NLtoLoc}) and Eq.(\ref{CorrectionFactor}) is given in Appendix \ref{Derivation}. 

In the asymptotic limit, the wave function takes the form

\begin{equation}\label{asymptotics}
\psi^{\textrm{asym}}_\ell(r)=\frac{i}{2}\left[H^-_\ell(\eta,kr)-\textbf{S}_{\ell j}H^+(\eta,kr) \right],
\end{equation}

\noindent where $\eta=Z_1Z_2e^2\mu/\hbar^2k$ is the Sommerfeld parameter, $k$ is the wave number, $\textbf{S}_{\ell j}$ is 
the scattering matrix element, and $H^-$ and $H^+$ are incoming and outgoing spherical Hankel functions, respectively. 
For neutrons, $\eta=0$.

In \Sec{Transfer}, we use the Distorted Wave Born Approximation (DWBA) to calculate the T-matrix  for the B$(p,d)$A reaction,
which, neglecting the remnant term, is written as

\begin{equation}\label{Tmatrix}
T_{p,d}=\langle\psi_{dA}^{(-)}\phi_d|V_{np}|\psi_{pB}\phi_{nA}\rangle \;,
\end{equation}

\noindent where $\psi_{dA}^{(-)}$ is the deuteron scattering wave function, $\phi_d$ is the deuteron bound state, $V_{np}$ is the 
Reid soft core $np$ interaction \cite{Reid1968_AP}, $\psi_{pB}$ is the proton distorted wave, and $\phi_{nA}$ is the neutron bound state wave function. 
(for details on the formalism, please check \cite{Thompson2009_Book}).

Due to its simplicity, a common technique is to do a calculation with a suitable local equivalent potential, then introduce 
the non-locality by modifying the wave function with the PCF

\begin{equation}\label{eq:pcf}
\psi^{PCF}_\ell(r)=F(r)\psi^{Loc}_\ell.
\end{equation}

\noindent This is precisely the approach we want to test in this study.

\section{Numerical details}
\label{numerical}

In this systematic study, we consider elastic scattering $(p,p)$ on $^{17}$O, $^{41}$Ca, $^{49}$Ca, $^{127}$Sn, $^{133}$Sn, 
and $^{209}$Pb at $20$ and $50$ MeV and the wave functions for a neutron bound to $^{16}$O, $^{40}$Ca, $^{48}$Ca, $^{126}$Sn, 
$^{132}$Sn, and $^{208}$Pb. In both cases, the full non-local equation is solved using the Perey-Buck potential, 
with the method described in Appendix \ref{SolvingEquation}.

For the scattering process, a local equivalent 
potential is determined by fitting the elastic scattering generated from the non-local equation. This was done using the 
code {\sc sfresco} \cite{fresco}. Using the local equivalent potential, the local scattering equation is solved to obtain $\psi^{Loc}$ and, 
finally, the PCF is applied to the wave function Eq.(\ref{eq:pcf}). The corrected wave function, $\psi^{PCF}$, is then compared to the solution 
of the full non-local equation, $\psi^{NL}$. 

A similar procedure is followed for the bound states. The full non-local equation is solved using a real Woods-Saxon form 
with radius $r=1.25$ fm, diffuseness $a=0.65$ fm, and a non-locality parameter $\beta=0.85$ fm. 
The depth is then adjusted to reproduce the physical binding energy of the system. The local equation is solved
with the local depth $V_{ws}^{Loc}$ necessary to reproduce the binding energy. We then apply the PCF to the resulting wave function, 
and renormalize to unity, to obtained the corrected bound state. The corrected wave function, $\phi^{PCF}$, is then compared to the solution 
of the full non-local equation, $\phi^{NL}$. 

The bound and scattering states resulting from either non-local or local potentials are then introduced into the DWBA 
T-matrix for $(p,d)$, Eq.(\ref{Tmatrix}), for describing the process at $20$ and $50$ MeV. Angular distributions are calculated using 
the code \textsc{fresco} \cite{fresco}. Non-locality was only added in the entrance channel, namely through the proton distorted wave and the neutron bound state. 
The local global parameterization of Daehnick \textit{et al.}  was used to obtain $\psi_{dA}$ in the exit channel. The scattering wave 
functions were solved by using a $0.05$ fm radial step size with a matching radius of $40$ fm. For the bound state solutions, 
we used a radial step size of $0.02$ fm. The matching radius was half the radius of the nucleus under consideration, and the 
maximum radius was $30$ fm, except for a very low binding energy study, when a larger value was necessary. 
The cross sections contain contributions of partial waves up to $J=30$.

In the following subsection, we present the results and analyze the effect of non-locality and the approximate correction 
factor in detail.

\begin{figure}[h]
\begin{center}
\includegraphics[width=0.5\textwidth]{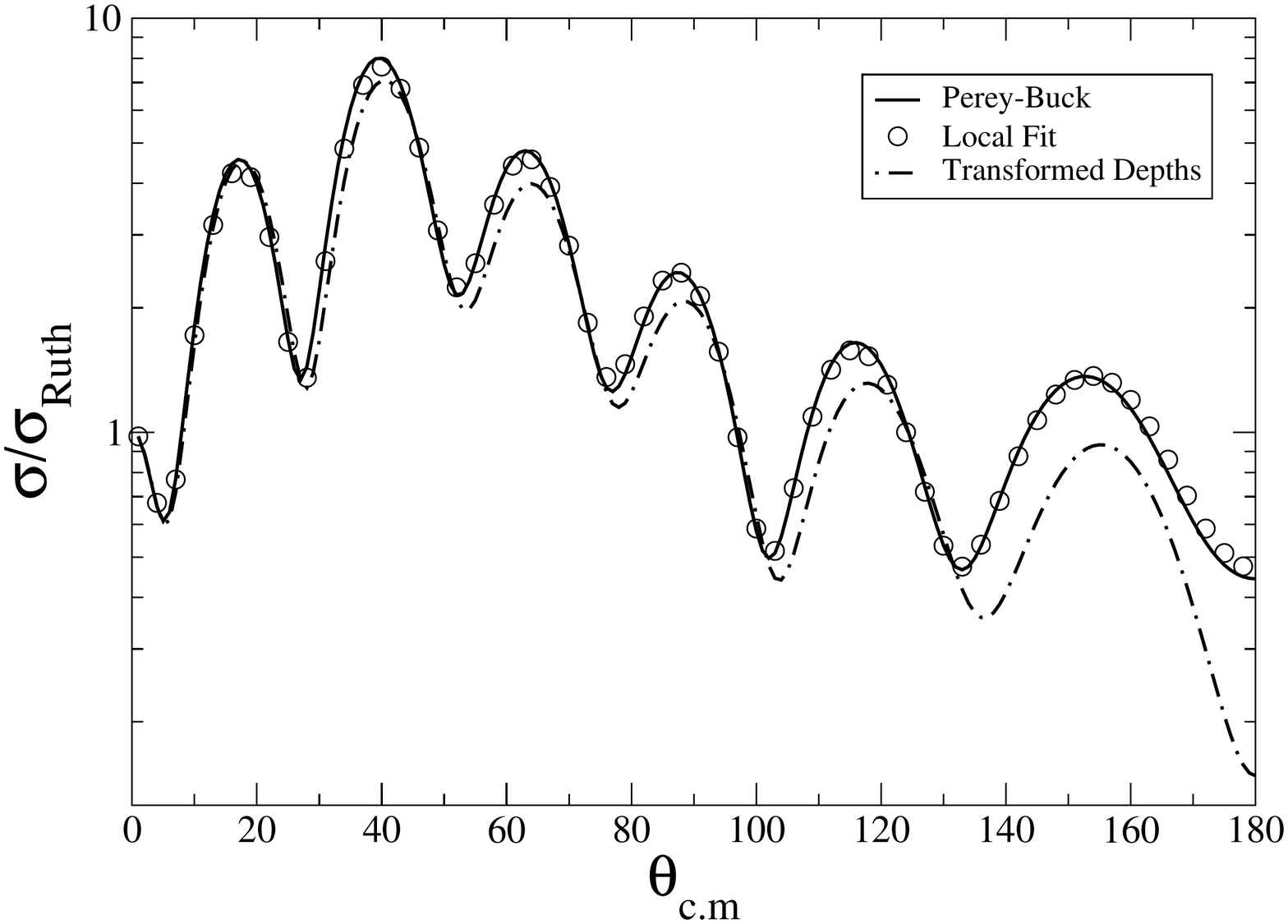}
\end{center}
\caption{$^{49}$Ca$(p,p)^{49}$Ca at $50.0$ MeV: the solid  line is obtained from solving Eq.(\ref{NLeqn}), the open circles 
are a fit to the non-local solution, and the dot-dash line is obtained by transforming the depths of the 
volume and surface potentials according to Eq.(\ref{NLtoLoc}).}
\label{fig:p49Ca_50-0_CS}
\end{figure}

\section{Results}
\label{results}


\subsection{Local Equivalent Potentials}
\label{LEP}

As described before, in order to study the correction factor, a local equivalent potential (LEP) needs to be found. A local 
potential is equivalent to a given non-local potential if it produces the same S-matrix elements, thus, producing the same 
elastic scattering  angular distribution. A LEP is found by $\chi^2$ minimization starting from the transformed local potential
obtained by using Eq.(\ref{NLtoLoc}). 
We required that the spin-orbit and Coulomb terms of the Perey-Buck non-local potential and the LEP be exactly the same, 
thus only the real volume and imaginary surface terms were allowed to vary in the fit to find the LEP (a total of 6 parameters). 
For most cases we were able to obtain a near perfect fit. We demonstrate the procedure with the elastic scattering of 
protons on $^{49}$Ca at 50 MeV. The ratio to Rutherford angular distributions are shown in Fig.\ref{fig:p49Ca_50-0_CS} 
for the case of the Perey-Buck non-local potential (solid line) and the LEP (circles). 
The angular distribution for the LEP sits on top of the one obtained from Perey-Buck, as it should.


\subsection{Transformed Local Equivalents}
\label{approximate_LEP}

In order to obtain the LEP, we first had to solve the non-local equation, which in the past was too demanding computationally.
Instead, the transformations shown in Eq.(\ref{NLtoLoc}) were sometimes used. We consider again the case of 
$^{49}$Ca$(p,p)^{49}$Ca at $50$ MeV. 
Table \ref{tab:Potential_Parameters} contains the original Perey-Buck potentials, the transformed potentials and the LEP. 
The subscripts $v$, $d$, $so$, and $c$ denote the real volume, imaginary surface, spin-orbit, and Coulomb terms of the potential, respectively. 
$V_v$, $W_v$ and $V_{so}$ are the depths of the potentials in MeV. The radius parameter, $r$, is used to find the radius of 
the nucleus under consideration through the formula $R=rA^{1/3}$, and $a$ is the diffuseness of the Woods-Saxon potential.

\begin{table}[h]
\centering
\begin{tabular}{|c|r|r|r|}
\hline
& Perey-Buck & Transformed & Local Fit  \\
\hline 
$V_{v}$ (MeV) & 71.000 & 37.151 & 37.842 \\ 
$r_{v}$ (fm) & 1.220 & 1.220 & 1.251 \\ 
$a_{v}$ (fm) & 0.650 & 0.650 & 0.629 \\ 
$W_{d}$ (MeV) & 15.000 & 7.849 & 8.697 \\ 
$r_{d}$ (fm) & 1.220 & 1.220 & 1.236 \\ 
$a_{d}$ (fm) & 0.470 & 0.470 & 0.440 \\ 
$V_{so}$ (MeV) & 7.180 & 7.180 & 7.180 \\
$r_{so}$ (fm) & 1.220 & 1.220 & 1.220 \\
$a_{so}$ (fm) & 0.650 & 0.650 & 0.650 \\
$r_{c}$ (fm) & 1.220 & 1.220 & 1.220 \\
\hline
\end{tabular}
\caption{Potential parameters for the reaction $^{49}$Ca$(p,p)^{49}$Ca at $ 50.0$ MeV: the first column are the parameters 
for the Perey-Buck non-local potential (the range of the non-locality in the Perey-Buck potentials is fixed $\beta=0.85$ fm), the second column corresponds to the local potential where the depths of the real volume and imaginary surface 
terms were transformed according to Eq.(\ref{NLtoLoc}), and the third column are the parameters for the local fit to the 
elastic distribution generated with the Perey-Buck potential. }
\label{tab:Potential_Parameters}
\end{table}

Table \ref{tab:Potential_Parameters} shows that the depths of the transformed local potential and the local fit are in 
fair agreement. In the local fit, the depth, radius, and diffuseness are adjusted for a better fit, while for the transformed 
local potential, the radius and diffuseness are the same as in the Perey-Buck potential. The dot-dashed line in Fig.
\ref{fig:p49Ca_50-0_CS} depicts the elastic scattering obtained from the transformed potential.  Aside from significant 
discrepancies at large angles, the transformed local potential does a fair job in reproducing the non-local elastic cross section. 
However, using this transformation formula to construct a potential should be treated with caution, since the fact that 
the radius and diffuseness are unchanged does not allow for a correct description of the diffraction pattern. 


\subsection{Corrections to Scattering Wave Functions}
\label{WFs}

Once the LEP solution $\psi^{Loc}(r)$ is found, the PCF can be tested by comparing it to the solution to the full non-local 
equation, $\psi^{NL}(r)$, and the corrected wave function, $\psi^{PCF}(r)$.
\begin{figure}[h]
\begin{center}
\includegraphics[width=0.5\textwidth]{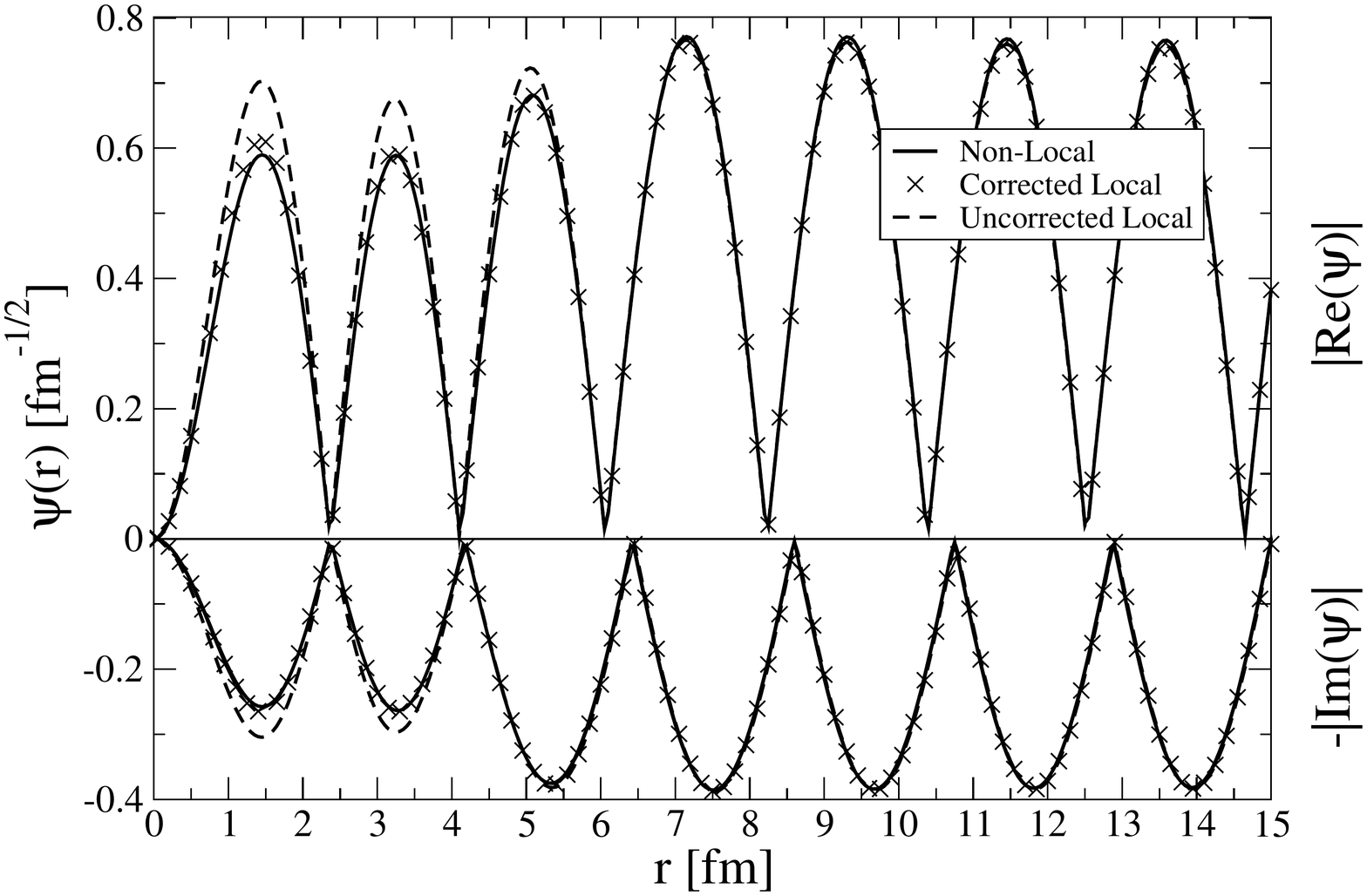}
\end{center}
\caption{The real and imaginary parts of the $J^\pi=0.5^-$ partial wave of the scattering wave function for the reaction 
$^{49}$Ca$(p,p)^{49}$Ca at $50.0$ MeV: $\psi^{NL}$ (solid line), $\psi^{PCF}$ (crosses) and $\psi^{Loc}$ (dashed line). Top (bottom) panel:  absolute value of the real (imaginary) part of the scattering wave 
function. }
\label{fig:p49Ca_50-0_001_001}
\end{figure}

\begin{figure}[h]
\begin{center}
\includegraphics[width=0.5\textwidth]{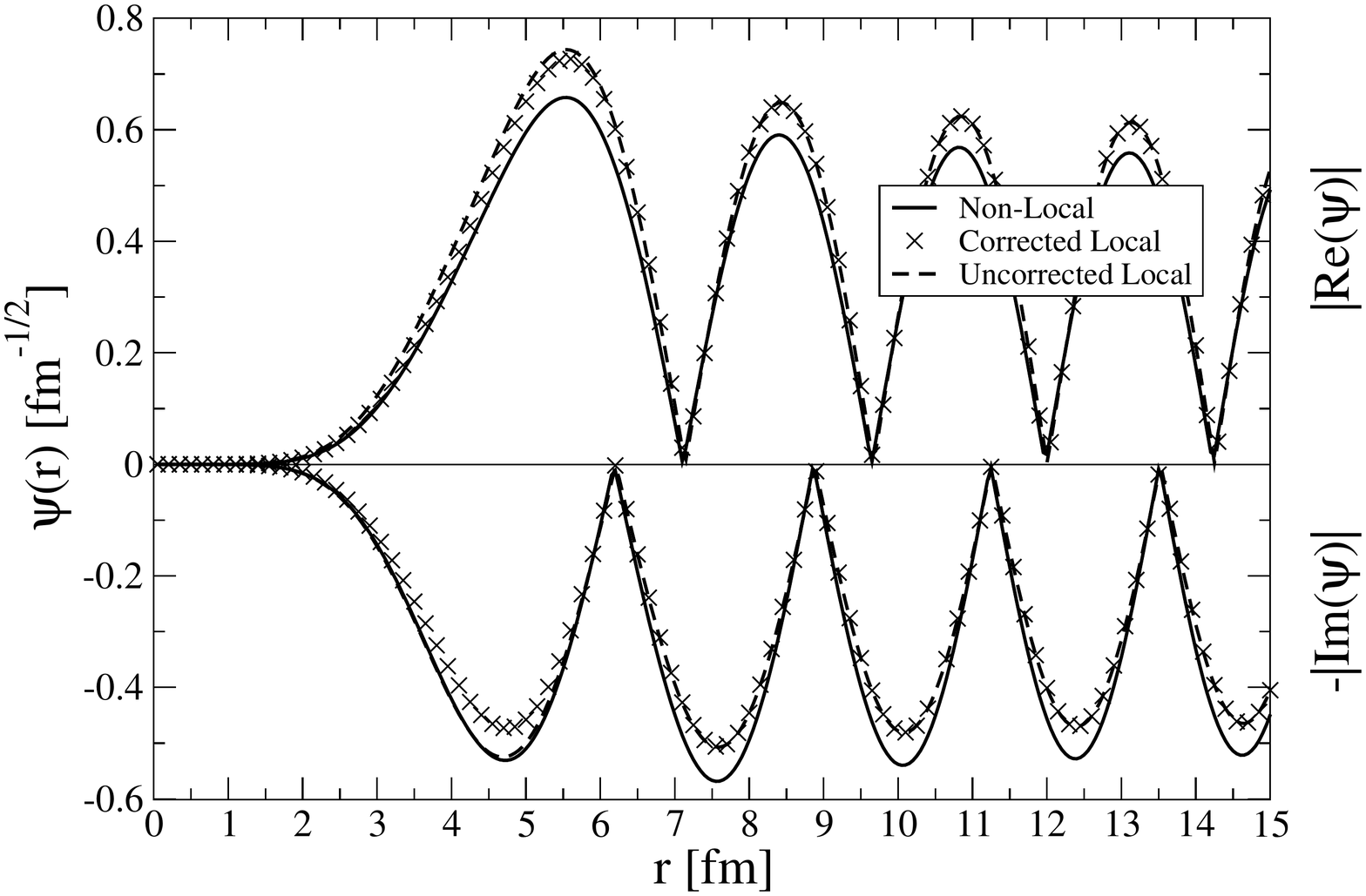}
\end{center}
\caption{The real and imaginary parts of the $J^\pi=5.5^+$ partial wave of the scattering wave function for the reaction 
$^{49}$Ca$(p,p)^{49}$Ca at $50.0$ MeV (see caption of Fig.\ref{fig:p49Ca_50-0_001_001}). }
\label{fig:p49Ca_50-0_006_011}
\end{figure}

For all cases investigated, the non-local wave function and the PCF corrected local wave function agree well for most 
partial waves. An example is provided in Fig.\ref{fig:p49Ca_50-0_001_001}. $\psi^{NL}$ (solid line) is reduced in the 
interior compared to $\psi^{Loc}$ (dashed line) but PCF accounts well for this reduction as shown by $\psi^{PCF}$ (crosses). 
However, in all cases, problems arose for partial waves corresponding to impact parameters around the surface region, 
as illustrated in Fig.\ref{fig:p49Ca_50-0_006_011}. The differences for these angular momenta are particularly relevant 
for transfer cross sections, which tend to be most sensitive to the surface region. The weaker performance of the PCF 
for partial waves corresponding to the surface is partly due to neglecting the $\nabla^2F$ term in the 
derivation of Eq.(\ref{CorrectionFactor}), which only contributes in the surface region (See Appendix \ref{Derivation}).

Occasionally we found slight differences in the asymptotic region  due to small differences in the S-Matrix elements 
for a particular partial wave. Since the wave functions are normalized according to Eq.(\ref{asymptotics}), small changes 
in the S-Matrix will result in different amplitudes for the real and imaginary parts of the scattering wave function in the 
asymptotic region. 


\subsection{Bound State Wave Functions}
\label{Bound}

For the bound state case, the non-local and local equivalent potentials were chosen so that the correct binding energy was 
reproduced. Non-local volume and local spin-orbit terms were included. Only the depth of the central volume potential was varied 
to reproduce the binding energy.  Since neutron bound states are of interest in $(d,p)$ and $(p,d)$ reactions, 
only these were considered. Like in the scattering case, Eq.(\ref{CorrectionFactor}) was used to correct the local wave function 
for non-locality. After the local wave function was corrected, the resulting wave function was renormalized. 
It is very important to renormalize the corrected wave function after applying Eq.(\ref{CorrectionFactor}). If the bound wave 
function is not normalized after applying the PCF, the resulting corrected wave function is worse than the 
uncorrected wave function.

\begin{figure}[h]
\begin{center}
\includegraphics[width=0.75\textwidth]{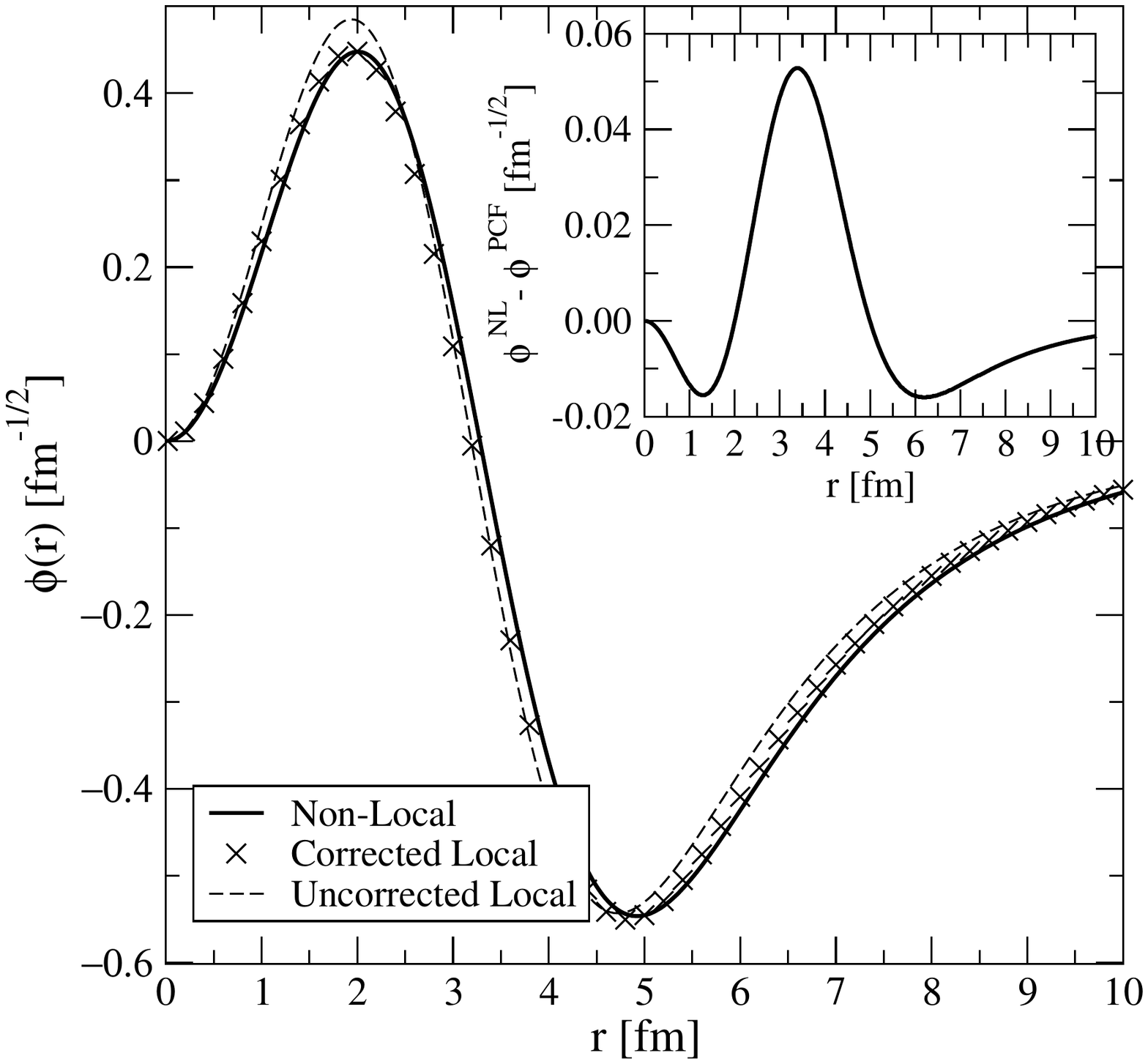}
\end{center}
\caption{The ground state, $2p_{3/2}$, bound wave function for $n+^{48}$Ca (see caption of Fig.\ref{fig:p49Ca_50-0_001_001}).
The inset shows the difference $\phi^{NL}-\phi^{PCF}$.}
\label{fig:n48Ca_Bound}
\end{figure}

The $2p_{3/2}$ ground state wave function for $n+^{48}$Ca is shown in Fig.\ref{fig:n48Ca_Bound}. Visually, the correction 
factor does an adequate job correcting for non-locality in the bound state. However, in the region between the two peaks of 
the wave function ($2-5$ fm), the PCF does very little to bring the local wave function into agreement with the non-local 
wave function. The inset in Fig.\ref{fig:n48Ca_Bound} shows the difference between $\phi^{NL}$ and $\phi^{PCF}$ as a function of r. 
The bound wave function has a large slope in this region, so the percent difference between the non-local and 
corrected local wave functions can be large. Also, the amplitude in the asymptotic region of the local wave functions are smaller 
than the amplitude of $\phi^{NL}$.  As we shall discuss next, in Section \ref{Transfer}, these features influence the $(p,d)$ 
transfer cross section in different ways.

\subsection{Transfer Cross Sections}
\label{Transfer}

\begin{figure}[h]
\begin{center}
\includegraphics[width=0.5\textwidth]{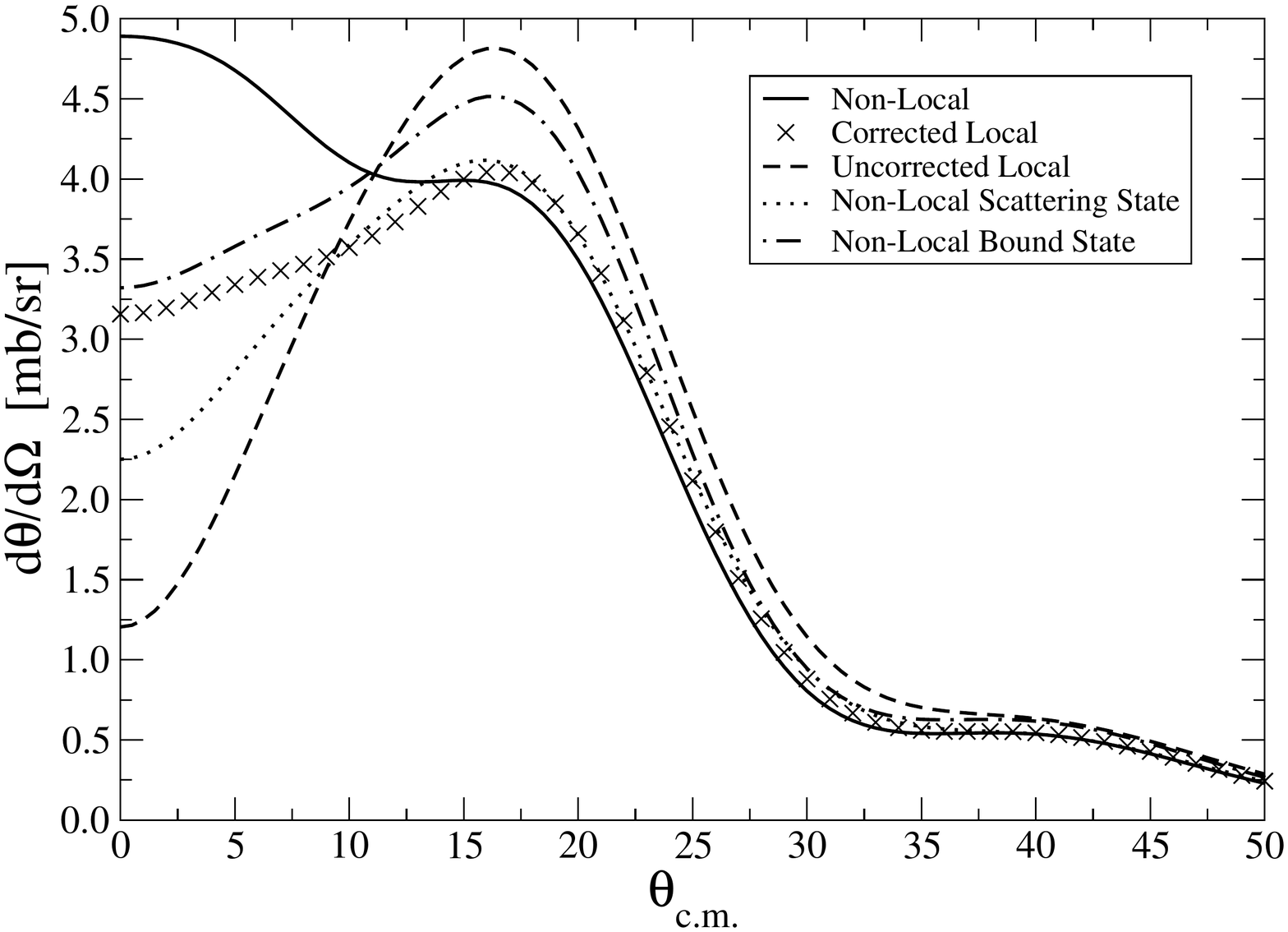}
\end{center}
\caption{Angular distributions for $^{49}$Ca$(p,d)^{48}$Ca at $50.0$ MeV: inclusion of non-locality in both the proton 
distorted wave and the neutron bound state (solid line), using LEP, then applying the 
correction factor to both the scattering and bound states (crosses),  using the LEP 
without applying any corrections (dashed line); including non-locality only to the proton distorted wave (dotted line),  and including non-locality in the neutron bound state  only (dot-dashed line).}
\label{fig:49Ca_50-0pd}
\end{figure}

We now turn our attention to the transfer $(p,d)$ cross sections.
In Fig.\ref{fig:49Ca_50-0pd} the $(p,d)$ transfer cross section for $^{49}$Ca$(p,d)^{48}$Ca at 50.0 MeV in the proton laboratory frame is shown. 
The solid line corresponds to including non-locality in both the proton distorted wave and the neutron bound state, the dashed 
line corresponds to the distribution obtained when only local equivalent interactions are used, and the crosses correspond 
to the cross sections obtained when the proton scattering state and the neutron bound state are both corrected by the PCF. 
While the Perey correction improves upon the distribution involving local interactions only, it is still unable to fully 
capture the complex effect of non-locality. The prominent changes at zero degrees  was unique to this case, but the very 
significant changes around the main peak was seen for most distributions studied. 

We also show the separate effect of including only non-locality in the proton scattering state (dotted) and the neutron 
bound state (dot-dashed). For this case, the non-locality in the proton distorted wave acts in a similar way to the non-locality 
in the bound state, namely it increases the cross section at zero degrees and reduces the cross section around $15^{\circ}$. 

The reason for the large changes at small angles can be seen from an analysis of the scattering and bound wave functions of 
Figs. \ref{fig:p49Ca_50-0_001_001}, \ref{fig:p49Ca_50-0_006_011}, and \ref{fig:n48Ca_Bound}. The existence of a node in the 
bound state wave function influences the cross section in a complex manner. The radius that corresponds to the surface for 
$^{49}$Ca occurs at a radius slightly larger than that where the bound state wave function is zero. 
The bound wave function has a large slope in this region, so the percent difference between the non-local and local wave 
functions can be quite large in this region. For this case, the non-local bound wave function is smaller than the local 
wave functions in this region, reducing the cross section at the peak. On the other hand, the magnitude of the bound wave function 
is larger for the non-local case in the tail region, which enhances the cross section at forward angles.

For the scattering wave functions, the largest differences were for partial waves that corresponded to the surface. Also, the 
asymptotics of scattering partial waves were different due to small differences in the S-Matrix, mostly for surface partial waves. 
The larger the amplitude in the asymptotic region, the larger the cross section at forward angles. There is an interplay between 
the real and imaginary parts of the scattering wave function which influences the cross section at forward angles. In a very 
complex manner, the combination of all these effects produces the interesting behavior of the transfer cross section at forward 
angles, and the changes in the magnitude of the cross section at the peak for this particular reaction.

In order to better understand this case, we artificially modified the bound wave function. By changing the binding energy 
we altered the Q-value of the reaction. Different Q-values produced very different types of distributions, both in shape 
and in magnitude. Nevertheless, similar dramatic changes in the cross section due to non-locality were found.
For very low binding energy, the normalization of the bound wave function was dominated by the asymptotics, so the PCF did very 
little. The node in the wave function altered the cross section in a very complex way. The PCF was not able to correct the 
bound wave function in the region around the node since the wave function and the PCF have a very large slope in this region, 
so inadequacies of the PCF were amplified. 

\begin{figure}[h]
\begin{center}
\includegraphics[width=0.5\textwidth]{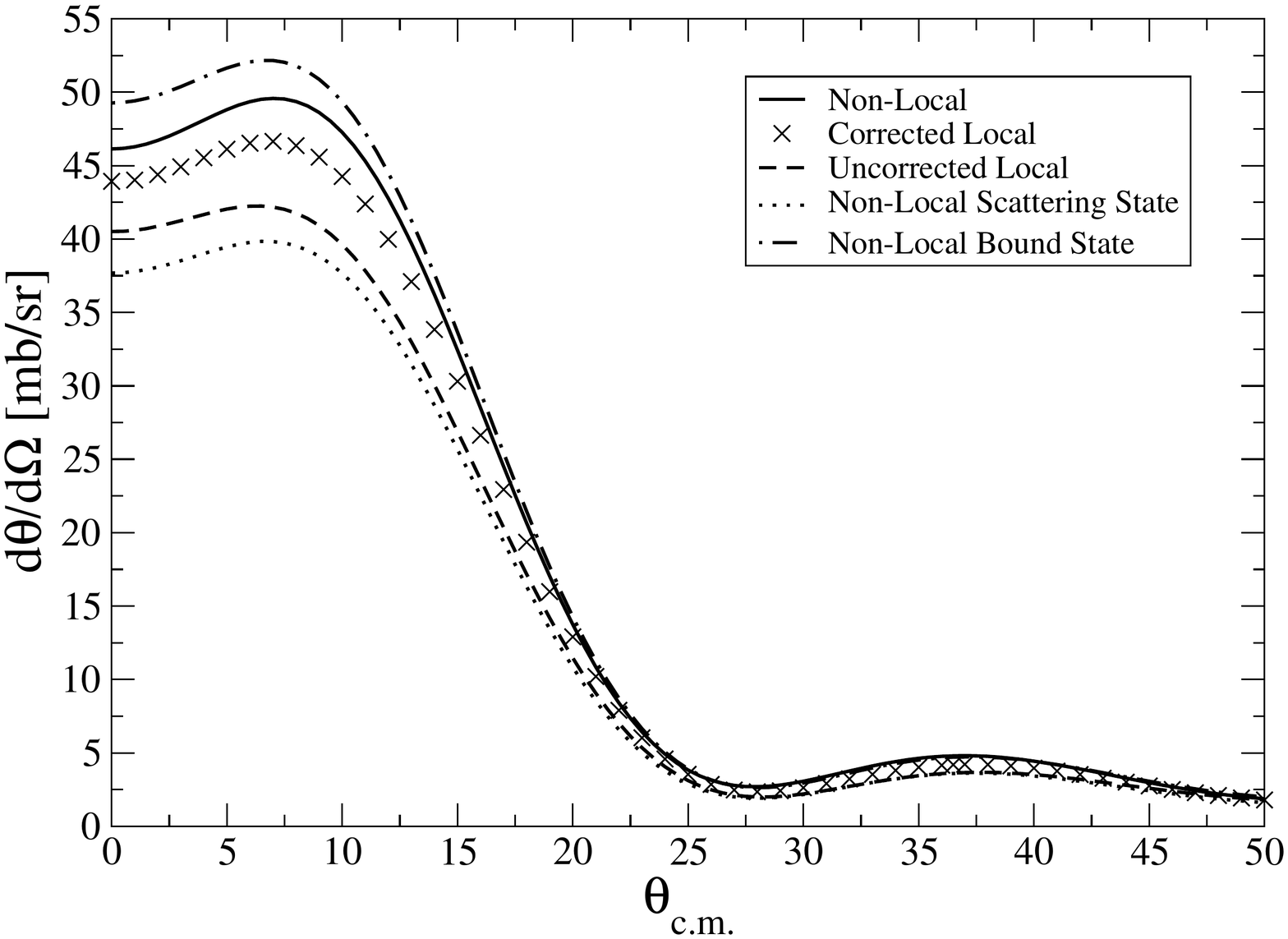}
\end{center}
\caption{Angular distributions for $^{49}$Ca$(p,d)^{48}$Ca at $20.0$ MeV (descriptions of each line is given in the 
caption of Fig.\ref{fig:49Ca_50-0pd}).}
\label{fig:49Ca_20-0pd}
\end{figure}

Consider now the same target but lower energy. In Fig.\ref{fig:49Ca_20-0pd} we present the transfer angular distribution for 
$^{49}$Ca$(p,d)^{48}$Ca at $20.0$ MeV. Non-locality is seen to have a large effect at small angles. Including non-locality in only 
the bound state increases the cross section at forward angles, which is to be expected from Fig.\ref{fig:n48Ca_Bound}, 
where it is seen that the magnitude of $\phi_{nA}^{NL}$ is larger than $\phi_{nA}^{PCF}$ and $\phi_{nA}^{Loc}$ in the asymptotic region. 
Non-locality in only the scattering state decreases the cross section, but only by a small 
amount. The net effect of non-locality is an overall increase in the cross section of 
$17.3\%$ relative to the  cross section obtained with local interactions only. While the correction factor moves the  transfer distribution in the right 
direction, it falls short by $5.2\%$.

Next we consider some heavier targets, $^{133}$Sn and $^{209}$Pb, and study $(p,d)$ at $20$ MeV. In both cases, the inclusion 
of non-locality in the scattering state decreases the cross section by a small amount. This is due to the low energy of the proton, 
and the high charge of the target; the details of the scattering wave function within the nuclear interior are not 
significant for the transfer since these details are suppressed by the Coulomb barrier. Non-locality in the bound state is very 
significant, and increases the cross section by a large amount in both cases. In $^{133}$Sn, the correction factor does a fair 
job taking non-locality into account, but there is still a noticeable discrepancy between the full non-local and corrected local 
results. In $^{209}$Pb, there are discrepancies at forward angles, but coincidentally the 
distributions resulting from the non-local potential and the local potential with the PCF
agree quite well at the major peak of the distribution. This agreement is accidental and comes from the non-local
effect in the bound state canceling that in the scattering state.

\begin{figure}[h]
\begin{center}
\includegraphics[width=0.5\textwidth]{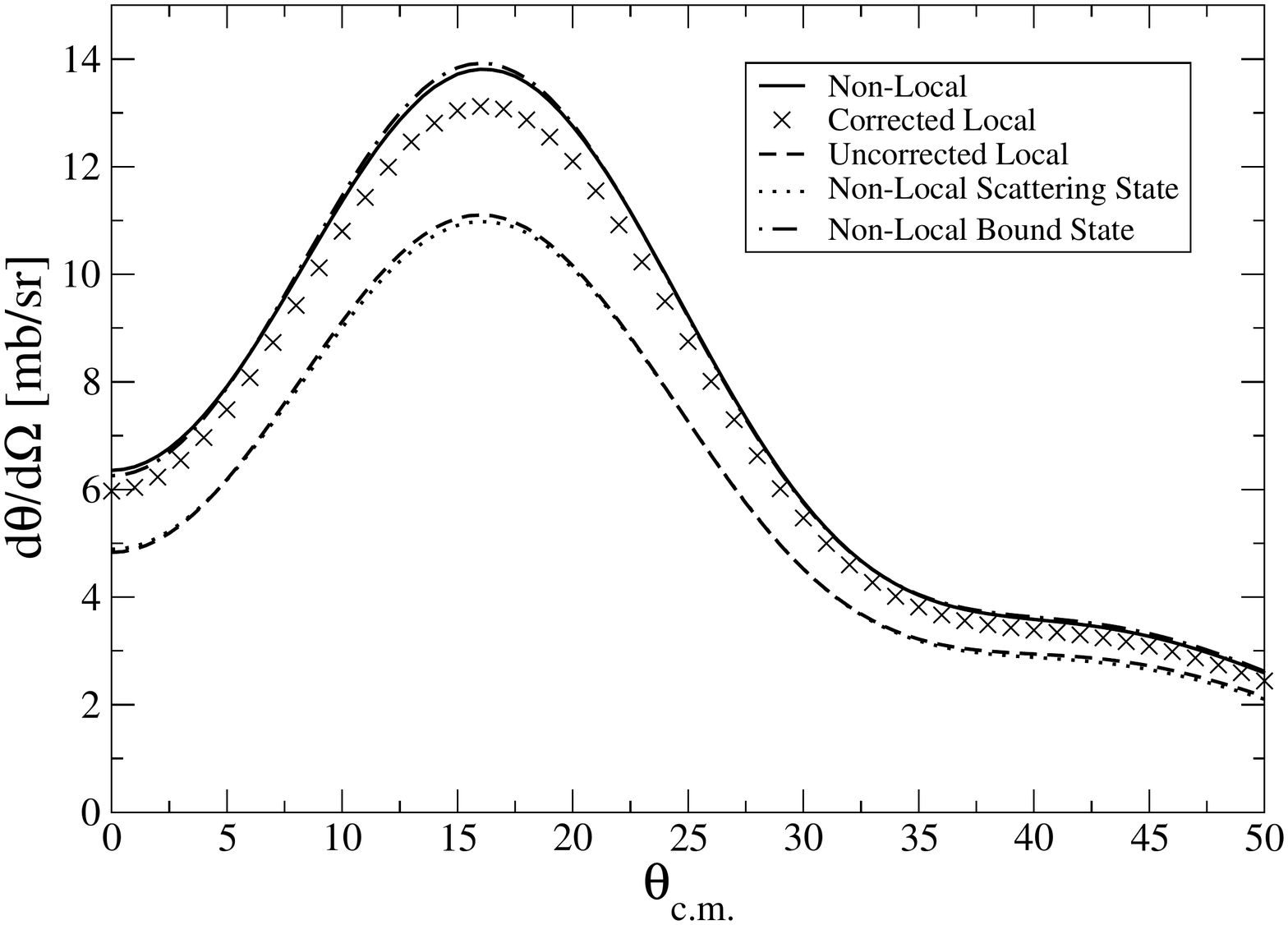}
\end{center}
\caption{Angular distributions for $^{133}$Sn$(p,d)^{132}$Sn at $20.0$ MeV (descriptions of each line is given in the 
caption of Fig.\ref{fig:49Ca_50-0pd}). }
\label{fig:133Sn_20-0pd}
\end{figure} 

\begin{figure}[h]
\begin{center}
\includegraphics[width=0.5\textwidth]{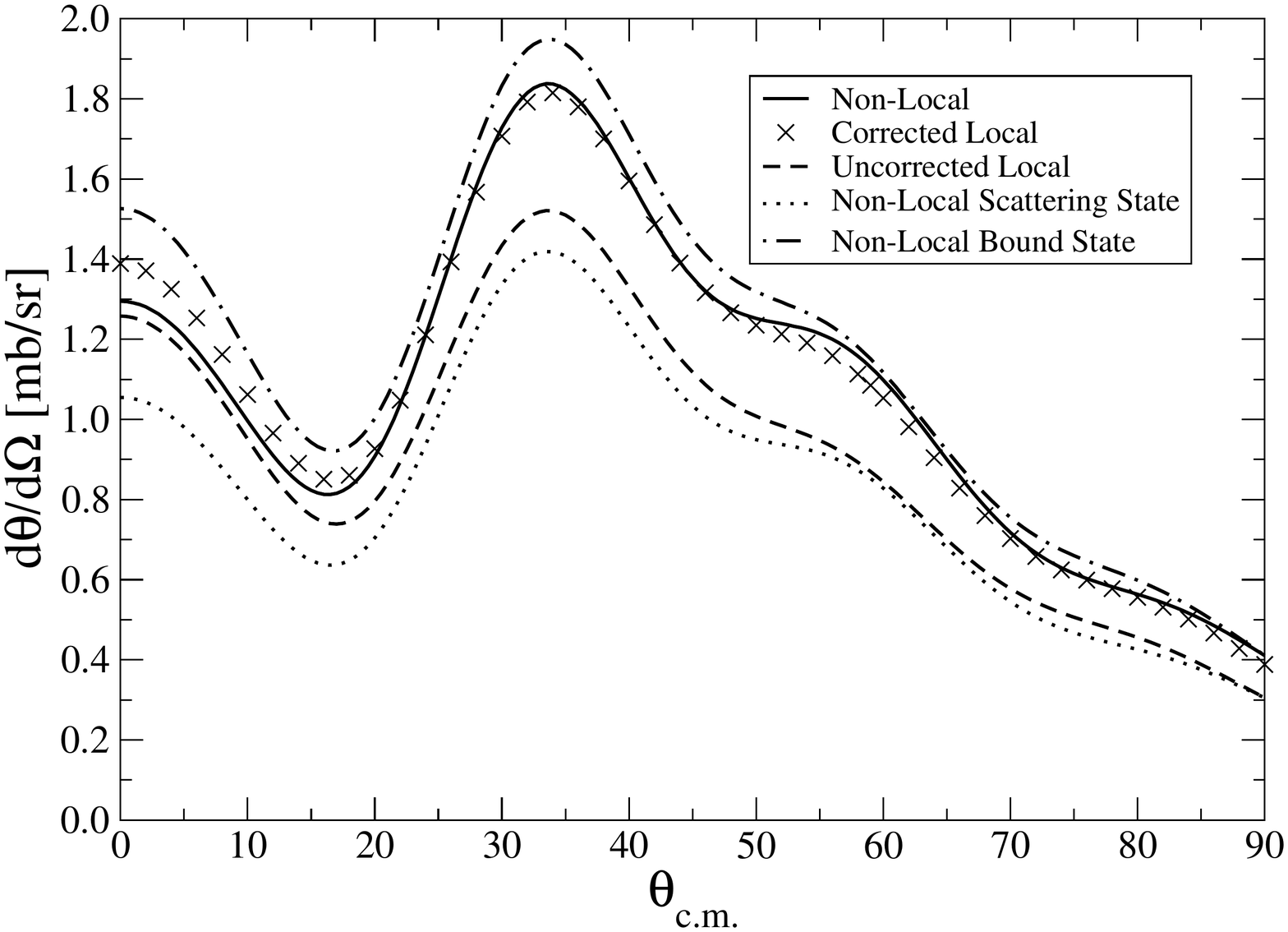}
\end{center}
\caption{Angular distributions for $^{209}$Pb$(p,d)^{208}$Pb at $20.0$ MeV (descriptions of each line is given in the 
caption of Fig.\ref{fig:49Ca_50-0pd}).}
\label{fig:209Pb_20-0pd}
\end{figure}


The percent differences at the first peak of the transfer distributions for all the cases that were studied are summarized 
in Table \ref{Tab:Percent_Difference_20} and \ref{Tab:Percent_Difference_50} for the $(p,d)$ reactions at 
$20$ and $50$ MeV.

\begin{table}[h]
\centering
\begin{tabular}{|c|r|r|}
\hline
& Corrected         & Non-Local          \\
$E_{lab}=20$ MeV & Relative to Local & Relative to Local  \\
\hline
$^{17}$O$(1d_{5/2})(p,d)$ & $7.1\%$  & $18.8\%$  \\ 
$^{17}$O$(2s_{1/2})(p,d)$ & $20.1\%$  & $26.5\%$ \\ 
$^{41}$Ca$(p,d)$ &$11.4\%$ & $21.9\%$     \\ 
$^{49}$Ca$(p,d)$ & $10.4\%$  & $17.3\%$   \\ 
$^{127}$Sn$(p,d)$  & $17.5\%$  & $17.3\%$ \\ 
$^{133}$Sn$(p,d)$ & $18.2\%$  & $24.4\%$  \\ 
$^{209}$Pb$(p,d)$ & $19.4\%$  & $20.8\%$  \\ 
\hline
\end{tabular}
\caption{Percent difference of the $(p,d)$ transfer cross sections at the first peak when using Eq.(\ref{CorrectionFactor}) (2nd column), or a non-local potential (3rd column), relative to the local calculation with the LEP, for a number of reactions occurring at 
$20$ MeV.}
\label{Tab:Percent_Difference_20}
\end{table}

\begin{table}[h]
\centering
\begin{tabular}{|c|r|r|}
\hline
& Corrected         & Non-Local  \\
$E_{lab}=50$ MeV & Relative to Local & Relative to Local \\
\hline
$^{17}$O$(1d_{5/2})(p,d)$  & $17.0\%$  & $35.4\%$  \\ 
$^{17}$O$(2s_{1/2})(p,d)$  & $0.2\%$  & $12.7\%$  \\ 
$^{41}$Ca$(p,d)$ & $2.9\%$  & $5.8\%$  \\ 
$^{49}$Ca$(p,d)$ & $-16.0\%$  & $-17.1\%$  \\ 
$^{127}$Sn$(p,d)$  & $10.1\%$  & $4.5\%$  \\  
$^{133}$Sn$(p,d)$ & $-6.7\%$  & $-16.9\%$  \\ 
$^{209}$Pb$(p,d)$ & $8.6\%$  & $8.6\%$  \\ 
\hline
\end{tabular}
\caption{Percent difference of the $(p,d)$ transfer cross sections at the first peak when using Eq.(\ref{CorrectionFactor}) (2nd column), or a non-local potential (3rd column), relative to the local calculation with the LEP, for a number of reactions occurring at 
$50$ MeV.}
\label{Tab:Percent_Difference_50}
\end{table}

It is seen that for both energies and for nearly all cases, the inclusion of non-locality in the entrance channel can have a 
very significant effect on the transfer cross section, often times introducing differences of $15-35\%$. 
Most of the time, adding non-locality increases the cross section at the first peak. In general, the correction factor moves 
the distribution obtained with local interactions  in the direction of the distribution including the non-local interactions. 
In the case of $^{127}$Sn$(p,d)$ at $50$ MeV, 
the correction factor overshoots  at the first peak, but the overall shape of the corrected 
distribution is in better agreement with the exact result. \\


\section{Discussion}
\label{discussion}

It should be noted that the PCF is only valid for non-local potentials of the Perey-Buck form. However, 
there is no reason to expect that the full non-locality in the optical potential will look anything like the Perey-Buck 
form. On physical grounds, the optical potential must be energy dependent due to non-localities arising from channel couplings. 
While the specific form chosen for the Perey-Buck potential is convenient for numerical calculations, a single Gaussian 
term mocking up all energy-independent non-local effects is likely to be an oversimplification. 

In an earlier study, Rawitscher \textit{et al.} \cite{Rawitscher1994_PRC} calculated the exchange non-locality in 
$n-^{16}$O scattering and examined the PCF. The wave functions obtained from their microscopically derived exchange 
non-locality were reasonably corrected by the PCF. The exchange non-locality is based on anti-symmetrized wave functions, 
which will naturally reduce the amplitude of the wave function in the nuclear interior due to the Pauli exclusion principle, 
similarly to the PCF. Results in \cite{Rawitscher1994_PRC} show that the PCF is able to approximately take into account the effects of including exchange.
In another study by Rawitscher \cite{Rawitscher1987_NPA}, the microscopic Feshbach optical potential from channel 
coupling is examined. The resulting potentials were strongly $\ell-$dependent, had emissive (positive imaginary) parts, 
and the non-local part did not resemble a Gaussian shape. The PCF obtained from the Wronskian was also strongly angular momentum 
dependent, and was found to be larger than unity in some cases. The channel coupling non-locality is therefore very different 
than the exchange non-locality, and one should not expect it to be corrected for in the same way. 
In those studies \cite{Rawitscher1994_PRC,Rawitscher1987_NPA}, the exchange and channel coupling non-localities were analyzed separately. 
To the authors knowledge, no study has examined the simultaneous inclusion of exchange and channel coupling non-localities. 

Our results, together with \cite{Rawitscher1994_PRC,Rawitscher1987_NPA}, emphasize the need for non-locality to be treated explicitly, contrary 
to what has been preferred  for more than 50 years. Since we have not yet found a good way to pin down non-locality phenomenologically, 
it would be extremely helpful to have microscopically derived optical potentials to guide further work. Microscopic $nA$ optical potentials based on 
the nucleon-nucleon interaction are particularly attractive because they immediately connect the intrinsic structure of the target to the reaction. 


\section{Conclusions}
\label{conclusions}

The long established Perey correction factor (PCF) was studied. To do so, the 
integro-differential equation containing the Perey-Buck non-local potential was solved numerically for single channel 
scattering and bound states. A local equivalent potential was obtained by fitting the elastic distribution generated by the 
Perey-Buck potential to a local potential. Both the local and non-local binding potentials reproduced the experimental 
binding energies. The scattering and bound state wave functions were used in a finite range DWBA calculation in order to 
calculate $(p,d)$ transfer cross sections. The PCF was applied to the wave functions generated with the 
local equivalent potentials. 

For the $(p,d)$ transfer reactions, we found that the explicit inclusion of non-locality to the entrance channel increased 
the transfer distribution at the first peak by $15-35\%$. The transfer distribution from using a non-local potential increased 
relative to the distribution from the local potential in most cases. In all cases, the PCF moved the transfer distribution 
in the direction of the distribution which included non-locality explicitly. However, non-locality was never fully taken into account with the PCF.


\begin{center}
\textbf{ACKNOWLEDGEMENT}
\end{center}

We are grateful to Jeff Tostevin for countless discussions and invaluable advice. We would also like to thank 
Nicolas Michel and Ron Johnson for many useful suggestions. This work was supported by the National Science
Foundation under Grant No. PHY-0800026 and the Department
of Energy under Contracts No. DE-FG52-08NA28552
and No. DE-SC0004087.


\appendix


\section{Deriving the Perey correction factor}
\label{Derivation}

Here we provide details on the derivation of the PCF, Eq.(\ref{CorrectionFactor}). We also include the derivation of the 
transformation formulas Eq.(\ref{NLtoLoc}), as well as the correct radial version of the transformation formulas which 
could be used to transform the non-local radius and diffuseness to their local counterpart.  

We start from Eq.(\ref{3D-NLeqn}). Let us define a function $F(\textbf{r})$ that connects the local 
wave function $\Psi^{Loc}(\textbf{r})$, resulting from the potential $U^{LE}(\textbf{r})$, with the wave function resulting
from a non-local potential, $\Psi^{NL}(\textbf{r})$

\begin{equation}{\label{locNLequivalence}}
\Psi^{NL}(\textbf{r}) \equiv F(\textbf{r})\Psi^{Loc}(\textbf{r}).
\end{equation}

Since the local and non-local equations describe the same elastic scattering, the wave functions should be identical 
outside the nuclear interior. Thus, $F(\textbf{r}) \rightarrow 1$ as $r \rightarrow \infty$. By inserting 
Eq.(\ref{locNLequivalence}) into the non-local equation Eq.(\ref{3D-NLeqn}) we can reduce the result to the following 
local equivalent equation

\begin{equation}
-\frac{\hbar^2}{2\mu}\nabla^2\Psi^{Loc}(\textbf{r})+U^{LE}(\textbf{r})\Psi^{Loc}(\textbf{r})=E\Psi^{Loc}(\textbf{r}),
\end{equation}

\noindent where the local equivalent potential is given by:

\begin{eqnarray}
U^{LE}(\textbf{r})&=&\frac{-\frac{\hbar^2}{\mu}\nabla F(\textbf{r})\cdot \nabla \Psi^{Loc}(\textbf{r})-\frac{\hbar^2}{2\mu}(\nabla^2F(\textbf{r}))\Psi^{Loc}(\textbf{r})}{F(\textbf{r})\Psi^{Loc}(\textbf{r})} \nonumber \\
&\phantom{=}&+\frac{\int U^{NL}(\textbf{r},\textbf{r}')F(\textbf{r}')\Psi^{Loc}(\textbf{r}')d\textbf{r}'}{{F(\textbf{r})\Psi^{Loc}(\textbf{r})}}+U_o(\textbf{r}). \nonumber \\
\end{eqnarray}

\noindent We next consider the second term of Eq.(A3) and introduce the explicit non-local potential form of Eq.(\ref{FrahnLemmer}). Using the definition 
$\textbf{s}=\textbf{r}-\textbf{r}'$,  expanding  in powers of $s$ up to first order, the integral becomes

\begin{eqnarray}
&\phantom{=}&\int U^{NL}_{WS}\left(\left|\textbf{r}-\frac{1}{2}\textbf{s}\right|\right)H(s)F(\textbf{r}-\textbf{s})\Psi^{Loc}(\textbf{r}-\textbf{s})d\textbf{s}  \nonumber \\
&\phantom{=}&\approx U_{WS}^{NL}(r)F(r)\int H(s)\Psi^{Loc}(\textbf{r}-\textbf{s})d\textbf{s} \nonumber \\
&\phantom{=}&-\frac{1}{2}F(r)\nabla U_{WS}^{NL}(\textbf{r})\cdot \int \textbf{s}H(s)\Psi^{Loc}(\textbf{r}-\textbf{s})d\textbf{s} \nonumber \\
&\phantom{=}&- \ U_{WS}^{NL}(r)\nabla F(\textbf{r})\int \textbf{s}H(s)\Psi^{Loc}(\textbf{r}-\textbf{s})d\textbf{s}
\end{eqnarray}

\noindent where

\begin{equation}
H(s)=\frac{\exp\left(-\frac{s^2}{\beta^2} \right)}{\pi^{3/2}\beta^3}.
\end{equation}

\noindent Therefore, the local equivalent potential becomes:

\begin{eqnarray}
U^{LE}(\textbf{r}) &\approx& \frac{1}{F(\textbf{r})\Psi^{Loc}(\textbf{r})}\left[ -\frac{\hbar^2}{\mu}(\nabla F \cdot \nabla \Psi^{Loc}) \right.  \nonumber \\
&\phantom{=}& + \ \left. U^{NL}_{WS}(r)F(r)\int H(s)\Psi^{Loc}(\textbf{r}-\textbf{s})d\textbf{s} \right. \nonumber \\
&\phantom{=}& - \ \left.  \frac{1}{2}F(r)\nabla U^{NL}_{WS} \cdot \int \textbf{s}H(s)\Psi^{Loc}(\textbf{r}-\textbf{s})d\textbf{s} \right. \nonumber \\
&\phantom{=}& - \ \left. U^{NL}_{WS}(r)\nabla F \cdot \int \textbf{s}H(s)\Psi^{Loc}(\textbf{r}-\textbf{s})d\textbf{s}\right] \nonumber \\
&\phantom{=}& - \ \frac{\hbar^2}{2\mu}\frac{\nabla^2 F(\textbf{r})}{F(\textbf{r})}+U_o(\textbf{r}).
\end{eqnarray}

\noindent Consider the four terms in the brackets. All of these terms are divided by $\Psi^{Loc}$, which has nodes. 
The first, third, and fourth terms depend on dot products and gradients of $\Psi^{Loc}$. These terms are unlikely 
to individually equal zero when $\Psi^{Loc}$ in the denominator equals zero. Thus, we require that these terms sum 
to zero so that $U^{LE}(\textbf{r})$ remains finite. As pointed out in \cite{Austern1970_Book}, this is not an approximation, but merely a condition for the method to work. 
Applying this condition gives us two equations:

\begin{eqnarray}\label{U_eqn}
U^{LE}(\textbf{r})&=&U^{NL}_{WS}(r)\left[\frac{\int H(s)\Psi^{Loc}(\textbf{r}-\textbf{s})d\textbf{s}}{\Psi^{Loc}(\textbf{r})} \right] \nonumber \\
&\phantom{=}&+U_o({\bf r})-\frac{\hbar^2}{2\mu}\frac{\nabla^2F(\textbf{r})}{F(\textbf{r})}
\end{eqnarray}

\begin{eqnarray}\label{0_eqn}
0&=&\frac{\hbar^2}{\mu}(\nabla F \cdot \nabla \Psi^{Loc})+\left[\frac{1}{2}F(r)\nabla U^{NL}_{WS}+U^{NL}_{WS}(r) \nabla F \right] \nonumber \\
&\phantom{=}&\cdot \int \textbf{s}H(s)\Psi^{Loc}(\textbf{r}-\textbf{s})d\textbf{s}.
\end{eqnarray}

\noindent Instead of using the local WKB approximations as in Austern \cite{Austern1970_Book}, we use the operator form of the Taylor expansion to factorize the wave function:

\begin{equation}\label{expansion}
\Psi^{Loc}(\textbf{r}-\textbf{s})=e^{-i\textbf{s}\cdot\textbf{k}}\Psi^{Loc}(\textbf{r}),
\end{equation}

\noindent with $\textbf{k}=-i\nabla$. This simplifies the integrals in Eq.(\ref{U_eqn}) and Eq.(\ref{0_eqn}). 
Consider first the integral in Eq.(\ref{U_eqn})

\begin{eqnarray}\label{integral}
\int H(s)\Psi^{Loc}(\textbf{r}&-&\textbf{s})d\textbf{s}=\left[\int e^{-i\textbf{s}\cdot\textbf{k}}H(s)d\textbf{s}\right]\Psi^{Loc}(\textbf{r}) \nonumber \\
&=& \exp\left[\frac{-k^2\beta^2}{4} \right]\Psi^{Loc}(\textbf{r}).
\end{eqnarray}

\noindent Therefore, assuming the potentials are scalar functions of $r$, and replacing Eq.(\ref{integral}) into Eq.(\ref{U_eqn}) we obtain;

\begin{eqnarray}\label{U_LE}
U^{LE}(r)&=&U^{NL}_{WS}(r)\exp\left[{-\frac{\mu\beta^2}{2\hbar^2}\left(E-U^{LE}(r) \right)}\right] \nonumber \\
&\phantom{=}&+U_o(r)-\frac{\hbar^2}{2\mu}\frac{\nabla^2F(r)}{F(r)},
\end{eqnarray}

\noindent where we used $k^2=-\nabla^2$ in the exponent to first order, and the Schr\"odinger's equation. Making the 
replacement $U^{LE}(r)=U^{Loc}_{WS}(r)+U_o(r)$, gives us the radial transformation formula

\begin{eqnarray}\label{radial}
U_{WS}^{NL}(r)&=&\left(U_{WS}^{Loc}(r)+\frac{\hbar^2}{2\mu}\frac{\nabla^2 F(r)}{F(r)} \right) \\
&\phantom{=}&\times \exp\left[\frac{\mu\beta^2}{2\hbar^2}\left(E-U_{WS}^{Loc}(r)-U_o(r) \right) \right]. \nonumber
\end{eqnarray}

\noindent The $\nabla^2F$ term is significant around the surface, but near the origin this term is negligible. 
Therefore, if we neglect this term, then we must remove the radial arguments, and consider this formula only near the origin. 
Therefore, for $r\approx0$

\begin{equation}\label{NLtoLoc3}
U_{WS}^{NL}(0)\approx U_{WS}^{Loc}(0)\exp\left[\frac{\mu\beta^2}{2\hbar^2}\left(E-U_{WS}^{Loc}(0)-U_o(0) \right) \right].
\end{equation}

The $U_{WS}(r)$ functions are of a Woods-Saxon form, and have real and imaginary parts

\begin{eqnarray}
U_{WS}(r)&=&U_R(r)+iU_I(r)  \\
&=&\frac{-V_v}{1+\exp\left(\frac{r-R}{a} \right)}+4i\frac{-W_d\exp\left(\frac{r-R}{a} \right)}{\left(1+\exp\left(\frac{r-R}{a} \right)\right)^2}. \nonumber
\end{eqnarray}

\noindent Inserting this into Eq.(\ref{NLtoLoc3}) we obtain;

\begin{eqnarray}\label{Trans}
&\phantom{=}&U_R^{NL}(r)+iU_I^{NL}(r)=(U_R^{Loc}(r)+iU_I^{Loc}(r)) \nonumber \\
&\times& \exp\left[\frac{\mu\beta^2}{2\hbar^2}\left(E-U_o(r)-U_R^{Loc}(r)-iU_I^{Loc}(r) \right) \right]. \nonumber \\
\end{eqnarray}

\noindent Near the origin, $U_{I}^{Loc}\approx 0$ so this term can be neglected in the exponent, and $U_R\approx-V_v$. 
While the spin-orbit term diverges at the origin, it rapidly goes to zero away from the origin, so we assume the spin-orbit 
contribution is negligible. Thus, $U_o=V_c$, where $V_c$ is the Coulomb potential at the origin for a uniform sphere of charge. 
Taking the real part of the above equation and making these substitutions gives

\begin{equation}
V^{NL}_v=V^{Loc}_v\exp\left[\frac{\mu\beta^2}{2\hbar^2}\left(E-V_c+V^{Loc}_v \right) \right],
\end{equation}

\noindent which is the first equation in Eq.(\ref{NLtoLoc}). For the imaginary part, we have:

\begin{equation}
U_I^{NL}(r)=U_I^{Loc}(r)\exp\left[\frac{\mu\beta^2}{2\hbar^2}\left(E-V_c+V^{Loc}_v \right) \right].
\end{equation}

\noindent While $U_I(r)\approx 0$ near the origin, the local and non-local terms have the same form factor, so the form factors 
exactly cancel as long as the radius and diffuseness are identical. Therefore, the imaginary part of Eq.(\ref{Trans}) gives

\begin{equation}
W^{NL}_d=W^{Loc}_d\exp\left[\frac{\mu\beta^2}{2\hbar^2}\left(E-V_c+V^{Loc}_v \right) \right],
\end{equation}

\noindent which is the second equation in Eq.(\ref{NLtoLoc}). It is important to note that these equations are only valid for 
transforming the depths of the potentials, thus Eq.(\ref{NLtoLoc3}) should not be used retaining the radial dependence. Indeed, Eq.(A13) is not valid for all r.

Now consider the integral in Eq.(\ref{0_eqn}). Using Eq.(\ref{expansion}) to expand the wave function, and evaluating the 
dot product we get

\begin{eqnarray}
0&=&\frac{\hbar^2}{\mu}(\nabla F \cdot \nabla \Psi^{Loc})+\left[\frac{1}{2}F(r)\nabla U^{NL}_{WS}+U^{NL}_{WS}(r) \nabla F \right] \nonumber \\
&\phantom{=}&\times \left[\int s\cos(\theta)H(s)e^{-i\textbf{s}\cdot\textbf{k}}d\textbf{s} \right]\Psi^{Loc}(\textbf{r}).
\end{eqnarray}

\noindent Doing the integral, we find that this becomes

\begin{eqnarray}\label{F-eqn}
0&=&\frac{\hbar^2}{\mu}\nabla F-\left[\frac{1}{2}F(r)(\nabla U^{NL}_{WS})+U^{NL}_{WS}(r)(\nabla F)\right] \nonumber \\
&\phantom{=}&\times\frac{\beta^2}{2}\exp\left[-\frac{\mu\beta^2}{2\hbar^2}\left(E-U^{LE}(r) \right) \right].
\end{eqnarray}

\noindent If we assume that the local momentum approximation is valid, this equation can be solved exactly and has the solution

\begin{equation}\label{CorrectionFactor2}
F(r)=\left[1-\frac{\mu \beta^2}{2 \hbar^2}U^{NL}_{WS}(r)\exp\left({-\frac{\mu\beta^2}{2\hbar^2}\left(E-U^{LE}(r) \right)}\right) \right]^{-\frac{1}{2}}.
\end{equation}

\noindent If the local momentum approximation is not valid, then insertion of Eq.(\ref{CorrectionFactor2}) into the $r.h.s.$ of Eq.(\ref{F-eqn}) 
will deviate from zero by a term related to the derivative of $U^{LE}(r)$. This additional term will be significant at the surface, and thus one 
can expect discrepancies in applying Eq.(\ref{CorrectionFactor2}) in this region. 

Comparing Eq.(\ref{CorrectionFactor2}) with Eq.(\ref{U_LE}) we see that

\begin{equation}
F(r)=\left[1-\frac{\mu \beta^2}{2\hbar^2}\left(U^{LE}(r)-U_o(r)+\frac{\hbar^2}{2\mu}\frac{\nabla^2F(r)}{F(r)} \right) \right]^{-\frac{1}{2}}.
\end{equation}

\noindent Neglecting the term containing $\nabla^2F$ gives us Eq.(\ref{CorrectionFactor}). The contribution of $\nabla^2F/F$ is only important at the surface, 
and again it is precisely for these radii that discrepancies can be expected in applying Eq.(\ref{CorrectionFactor}).


\section{Solving the Equation}
\label{SolvingEquation}

In order to assess the validity of the local approximation we need to solve Eq.(\ref{NLeqn}) exactly. For the scattering state, 
our approach follows Perey and Buck \cite{Perey1962_NP}, where Eq.(\ref{NLeqn}) is solved by iteration. For simplicity, we will 
drop the local part of the non-local potential, $U_o(r)$, in our discussion, although it is included in our calculations.

Scattering solutions are considered first, where the subscript $n$ denotes the $n$th order approximation to the correct solution. 
The iteration scheme starts with an initialization:

\begin{equation}\label{NLeqn2}
\frac{\hbar^2}{2\mu}\left[\frac{d^2}{dr^2}-\frac{\ell(\ell+1)}{r^2}\right]\psi_{n=0}(r)+[E-U_{\textrm{init}}(r)]\psi_{n=0}(r)=0,
\end{equation}

\noindent where $U_{\textrm{init}}(r)$ is some suitable local potential used to get the iteration process started. Knowing $\psi_o(r)$ one then proceeds with solving:

\begin{eqnarray}\label{NLeqn3}
&\phantom{=}&\frac{\hbar^2}{2\mu}\left[\frac{d^2}{dr^2}-\frac{\ell(\ell+1)}{r^2} \right]\psi_n(r)+[E-U_{\textrm{init}}(r)]\psi_n(r) \nonumber \\
&=&\int g_\ell(r,r')\psi_{n-1}(r') dr' -U_{\textrm{init}}(r)\psi_{n-1}(r),
\end{eqnarray}

\noindent with as many iterations necessary for convergence. The number of iterations for convergence depends mostly on the 
partial wave being solved for (lower partial waves require more iterations) and the quality of $U_{\textrm{init}}(r)$. It was 
rare for any partial wave to require more than 20 iterations to converge, even with a very poor choice for $U_{\textrm{init}}(r)$. 
If the LEP is used as $U_{\textrm{init}}(r)$, then any partial wave converges with less than 10 iterations.

For the bound state problem, the method is somewhat different. A variety of methods exist in the literature, some developed specifically to
 handle non-analytic forms (e.g. \cite{Michel2009_EPJ}). Our approach may not be the most efficient, but it is straightforward, general and easy to implement. 
To solve the bound state problem with a non-local potential we begin by solving Eq.(\ref{NLeqn2}). Since we are using the wave function from the 
previous iteration to calculate the non-local integral, we need to keep track of the different normalizations of the inward 
and outward wave functions that results from the choice for the initial conditions for each wave function. Thus, the 
equations we iterate are:

\begin{eqnarray}\label{Inward}
&\phantom{=}&\frac{\hbar^2}{2\mu}\left[\frac{d^2}{dr^2}-\frac{\ell(\ell+1)}{r^2} \right]\phi_n^{In}(r)+[E-U_{\textrm{init}}(r)]\phi_n^{In}(r) \nonumber \\
&=&\int_0^{R_{Max}} g_\ell(r,r')\phi_{n-1}^{In}(r') dr' -U_{\textrm{init}}(r)\phi_{n-1}^{In}(r)
\end{eqnarray}

\begin{eqnarray}\label{Outward}
&\phantom{=}&\frac{\hbar^2}{2\mu}\left[\frac{d^2}{dr^2}-\frac{\ell(\ell+1)}{r^2} \right]\phi_n^{Out}(r)+[E-U_{\textrm{init}}(r)]\phi_n^{Out}(r) \nonumber \\
&=&\int_0^{R_{Max}} g_\ell(r,r')\phi_{n-1}^{Out}(r') dr' -U_{\textrm{init}}(r)\phi_{n-1}^{Out}(r),
\end{eqnarray}

\noindent where $R_{Max}$ is some maximum radius chosen greater than the range of the nuclear interaction. Note that $\phi^{In}(r)$ is 
the wave function for integrating from the edge of the box inward and has a normalization set by  the Whittaker function 
as  initial condition, while $\phi^{Out}(r)$ is the wave function for integrating from the origin outward and has the 
normalization set  using the standard $r^{L+1}$ initial condition near the origin. 

Even though $\phi^{Out}$ and $\phi^{In}$ differ by only a constant, these two equations, Eq.(B3) and Eq.(B4), are necessary because the value 
of the normalization constant is only known after convergence. For a given iteration, $\phi^{Out}$ and $\phi^{In}$ converge 
when their logarithmic derivatives agree at the matching point. To keep the proper normalization throughout the entire 
range $[0,R_{Max}]$, we need to retain two versions of the converged wave function 
for each iteration:

\begin{equation}
 \phi_n^{\textrm{In}}(r) = \left\{
  \begin{array}{l l}
    C_n^{\textrm{Out}}\phi_n^{\textrm{Out}}(r) & \quad \textrm{for $0 \le r < R_{\textrm{Match}}$}\\
    \phi_n^{\textrm{In}}(r) & \quad \textrm{for $R_{\textrm{Match}} \le r \le R_{\textrm{Max}}$}
  \end{array} \right.
\end{equation}

\begin{equation}
 \phi_n^{\textrm{Out}}(r) = \left\{
  \begin{array}{l l}
    \phi_n^{\textrm{Out}}(r) & \quad \textrm{for $0 \le r < R_{\textrm{Match}}$}\\
    C_n^{\textrm{In}}\phi_n^{\textrm{In}}(r) & \quad \textrm{for $R_{\textrm{Match}} \le r \le R_{\textrm{Max}}$}
  \end{array} \right.
\end{equation}

\noindent where 

\begin{equation}
C^{\textrm{In(Out)}}=\frac{\phi^{\textrm{Out(In)}}(R_{\textrm{Match}})}{\phi^{\textrm{In(Out)}}(R_{\textrm{Match}})}
\end{equation}

The full iteration scheme is converged when the binding energy obtained from the previous iteration agrees with the 
binding energy from the current iteration within a desired level of accuracy. Although this may not be the most efficient method, it is general 
(whatever the form of non-locality) and is very stable, providing a good option for future studies beyond the Perey-Buck potentials.


\end{document}